\documentclass[peerreview,onecolumn,11pt,draftclsnofoot]{IEEEtran}

\usepackage{amsmath}
\usepackage{amsfonts}
\usepackage{epsfig}
\usepackage{amssymb}
\usepackage{cite}
\usepackage{subfigure}
\usepackage{multirow}
\usepackage{rotating}
\usepackage{graphicx}
\begin{document}

\title{Opportunistic Power Control for Multi-Carrier Interference Channels
\thanks{Manuscript received April 18, 2011. This work was supported in part by Tarbiat Modares University, and in part by Iran Telecommunications Research Center , Tehran, Iran under PhD Research Grant 89-09-95.}}
\author{Mohammad~R.~Javan and Ahmad~R.~Sharafat, ~\IEEEmembership{Senior Member,~IEEE}
\thanks{The authors are with the Department of Electrical and Computer Engineering, Tarbiat Modares University, P.~O.~Box 14155-4838, Tehran, Iran. Corresponding author is A. R. Sharafat (email: sharafat@modares.ac.ir).}}

%

\maketitle

\begin{abstract}
We propose a new method for opportunistic power control in
multi-carrier interference channels for delay-tolerant data services. In doing so, we utilize a game theoretic framework with novel constraints, where each user tries to maximize its utility in a distributed and opportunistic manner, while satisfying the game's constraints by adapting its transmit power to its channel. In this scheme, users transmit with more power on good sub-channels and do the opposite on bad sub-channels. In this way, in addition to the allocated power on each sub-channel, the total power of all users also depends on channel conditions. Since each user's power level depends on power levels of other users, the game belongs to the \emph{generalized} Nash equilibrium (GNE) problems, which in general, is hard to analyze. We show that the proposed game has a GNE, and derive the sufficient conditions for its uniqueness. Besides, we propose a new pricing scheme for maximizing each user's throughput in an opportunistic manner under its total power constraint; and provide the sufficient conditions for the algorithm's convergence and its GNE's uniqueness. Simulations confirm that our proposed scheme yields a higher throughput for each user and/or has a significantly improved efficiency as compared to other existing opportunistic methods.
\end{abstract}


\section{Introduction}\label{introduction}

Frequency spectrum of a network can be divided into many
orthogonal sub-channels that can be shared by all users. However,
shared usage of spectrum by a user produces interference to other
users. As the number of users increases, users' received
interference levels increase as well, resulting in fewer users
that can achieve their required quality of service (QoS). In such
instances, efficient use of spectrum becomes more important,
meaning that users should transmit at their lowest possible power
levels while satisfying their required data rates.

Radio resource allocation in multi-carrier wireless networks aims
to allocate available resources to users in such a way that under
some system and service constraints, a performance measure for
each user, e.g., the total throughput or the total transmit power, is optimized. In distributed resource allocation, each user
chooses its strategy (required resources) independent of other
users. Game theory \cite{fudenberg1} is commonly used in the
literature to analyze distributed resource allocation algorithms
by formulating the problem as a non-cooperative game, where each
user competes with other users with a view to optimizing its own
utility. The game settles at its Nash equilibrium (NE), where no
user can improve its utility by unilaterally changing its
strategy.

In the context of distributed resource allocation, the problem of
power minimization under throughput constraint is considered in
\cite{han1,scutari1}. In this problem, each user tries to minimize its total consumed power over all sub-channels while maintaining
its total throughput above a predefined threshold. In such a game, since the strategy space of each user depends on the chosen
strategies of other users, the game cannot be analyzed by way of
the conventional Nash equilibrium, and the so called
\emph{generalized Nash equilibrium} (GNE) should be used
\cite{facchinei1}. In \cite{han1}, the game is simulated, and in
\cite{scutari1}, the game is mathematically analyzed and the
sufficient conditions for the existence and uniqueness of GNE, as
well as the sufficient conditions for the convergence of the
distributed algorithm are presented.

The counterpart of this problem in single carrier systems is
known as the target SINR-tracking power-control algorithm (TPC)
\cite{foschini1}. In TPC, each user tries to choose its transmit
power in a distributed manner with a view to satisfying its
predefined target SINR. An important issue in TPC is its
feasibility, meaning that the algorithm converges only if a power
vector exists such that the target SINRs of all users are
satisfied. In TPC, users adapt their transmit power levels to
their channel conditions, i.e., each user increases its transmit
power when its channel is bad, and does the reverse when its
channel is good. Likewise, minimizing the transmit power under the data rate constraint may not be feasible, as there may not be a
power vector that satisfies the data rate constraints for all
users, meaning that no GNE may exist. Also, since users consume
more power when their channels are bad, they increase their
interference to others, which in turn forces others to increase
their transmit power levels, and thus aggravating the situation
further for all users.

To avoid such a case for delay-tolerant services, it is worthwhile for users to reduce their transmit power (which would result in
lower data rates) when their channels are bad, and do the reverse
when their channels are good. This is the opportunistic power
control (OPC) that was introduced and analyzed in \cite{sung1} for single carrier systems. The alternative OPC for multi-carrier systems is that users try to maximize their total throughput over all sub-channels in a distributed manner when the total transmit power for each user is constrained. This game has been extensively analyzed in \cite{scutari2,scutari3,scutari4,pang1,sung2}, and different conditions for the uniqueness of NE and for the convergence of the distributed algorithms are obtained. The drawbacks of this scheme include causing interference without any gain by a transmitting user when its channel is bad, and convergence to an inefficient NE because of self-serving and independent users.

In \cite{scutari2}, the data rate maximization game is analyzed
and the sufficient conditions for NE's uniqueness are obtained. In addition, by interpreting the waterfilling solution as a projector and using its contraction property \cite{bertsekas1}, in
\cite{scutari3,scutari4}, the sufficient conditions for the
convergence of the iterative waterfilling solution are derived.
Using linear complementarity \cite{pang1} reformulation of the
Karush-Kuhn-Tucker (KKT) condition \cite{boyd1} of the rate
maximization game, in \cite{pang2}, the linear complementarity
conditions are converted to the affine variational inequality
\cite{pang3}, and the sufficient conditions for NE's uniqueness
and convergence of the iterative waterfilling algorithm are
obtained. In \cite{sung2}, the sufficient conditions for NE's
uniqueness and convergence of the iterative waterfilling algorithm are established by interpreting the waterfilling solution as a
piecewise affine function \cite{pang3}.

In a non-cooperative game, since each user selfishly tries to
optimize its own utility, the equilibrium of the game may not be a desirable one. In such cases, pricing is an effective mechanism to control users' behaviors and achieve a more efficient NE. The
proposed pricing in \cite{krunz1,scutari5} for the rate
maximization game is a linear function of the transmit power. In
\cite{krunz1}, only numerical results are presented that provide
some insight on NE's uniqueness and convergence of the distributed algorithm, whereas in \cite{scutari5}, a mathematical analysis
based on the notion of variational inequalities and non-linear
complementarity is presented to obtain the sufficient conditions
for NE's existence and for convergence of the distributed
algorithm.

In this paper, we propose an OPC for multi-carrier interference channels using a game theoretic framework. In our scheme, each user opportunistically tries to maximize its total transmit power over all sub-channels in a distributed manner with a view to maximizing its data rate while satisfying the given constraint on the total transmit power that depends on interference levels in sub-channels. In doing so, higher transmit power levels are allocated to sub-channels with low interference, and lower power levels to high interference sub-channels. This is in contrast to the rate maximization game with total power constraint irrespective of interference levels on sub-channels \cite{han1, pang2}.

For the proposed game, we analyze the existence of GNE. Similar to the power minimization game with data rate constraint, the
strategy space of a user in our proposed scheme depends on the
strategies chosen by other users. However, we will show that at
least one GNE is guaranteed to exist for the proposed scheme. In
addition, we obtain the sufficient conditions for GNE's
uniqueness, and for the convergence of the distributed algorithm.
We also introduce pricing to the rate maximization game for
controlling selfish users and providing incentives for behaving in an opportunistic manner. In doing so, pricing is a function of the user's transmit power and the interference experienced by that
user. Furthermore, we obtain the sufficient conditions for NE's
uniqueness, and for the the distributed algorithm's convergence by utilizing variational inequalities as in
\cite{pang2,scutari1,scutari3,scutari5}. By way of simulation, we
evaluate the performance of our proposed OPC, as well as that of the pricing-based data rate maximization problem, and show that the total throughput of users is increased as compared to when pricing is not applied.

This paper is organized as follows. A brief review of the TPC and
the OPC algorithms is provided in Section \ref{tpcopc}. The
opportunistic power control problem is studied in Section
\ref{opcg}, and our pricing is introduced in Section
\ref{opcpricing}. Simulation results are presented in Section
\ref{simulation}, and conclusions are in Section \ref{conclusion}.

\section{TPC and OPC in Single Carrier Systems}\label{tpcopc}

Consider a single carrier wireless network with $M$ active users
that are spread in its coverage area. Let the channel gain between the transmitter of user $j$ and the receiver of user $i$ be
$G_{i,j}$, and noise power at the receiver of user $i$ be
$\eta_i$. When user $i$ transmits at power level $p_i$, its SINR,
denoted by $\gamma_i$ is
\begin{equation}\label{SINR}
    \gamma_{i} = \frac{{G_{i,i}}{p_i}}{\sum\limits_{j \neq
    i}{{G_{i,j}}{p_j}}+\eta_i}.
\end{equation}
We denote the effective interference by
\begin{equation}\label{rieff}
    I_{i}=\frac{\sum\limits_{j \neq i}{{G_{i,j}}{p_j}}+\eta_i}{G_{i,i}}.
\end{equation}
The value of $I_i$ depends on power levels of users (i.e.,
$I_i(\textbf{p})$), but for simplicity in notation, we use $I_i$.

Consider a case in which each user chooses its power level in a
distributed and iterative manner with a view to maintaining its
SINR above a predefined threshold, i.e., $\gamma_i \geq
\hat\gamma_i$. In this case, a distributed algorithm for achieving a given fixed target SINRs (TPC) as in \cite{foschini1} is
\begin{equation}\label{linearprogram}
    p_i(n+1)= \frac{\hat\gamma_i}{\gamma_i(n)}p_i(n).
\end{equation}
When the system is feasible, i.e., if there exists a power vector
$\textbf{p}=[p_1,\cdots,p_M]$ such that the SINR constraints are
satisfied, this algorithm converges to the solution of the
following problem
\begin{eqnarray}\label{linearprogram}
 \text{min} \sum p_i \\\nonumber
& & \hspace{-3.1cm} \text{subject to:}~\gamma_i \geq \hat\gamma ~
\qquad \forall i,
\end{eqnarray}
where all constraints are satisfied with the equality.

To maintain a predefined SINR, a user with a bad channel transmits at high power and causes interference to other users with no
apparent benefit to itself. To increase the system throughput, for delay-tolerant services, it is better that users with bad channels reduce their transmit power even to 0, and users with good
channels do the reverse, and both groups adapt their data rates to their respective transmit power levels. This is the OPC algorithm, defined by
\begin{equation}\label{OPC}
    p_i(n+1)= \frac{\zeta_i}{I_i(n)},
\end{equation}
where $\zeta_i$ is a predefined constant and $I_i(n)$ is the
effective interference experienced by user $i$ in iteration $n$ as defined in (\ref{rieff}). As such, each user transmits at a rate
given by
\begin{equation}\label{rate}
    R_i= \log(1+\gamma_{i}),
\end{equation}
where $\gamma_{i}$ is defined in (\ref{SINR}). From (\ref{OPC}),
it is clear that when the channel is bad, i.e., when $I_i$ is
high, the user decreases its transmit power, and when the channel
is good, i.e., when $I_i$ is low, it does the opposite. For a
given transmit power level, the OPC's throughput is higher than
that of TPC.

\section{Opportunistic Power Control}\label{opcg}

\subsection{Problem Formulation}\label{problem formulation}
We now formulate the opportunistic power control problem for
multi-carrier interference channels via a game theoretic
framework. We begin by considering a special power control problem
that is the TPC's counterpart in multi-carrier systems. In this
problem, each user tries to minimize its total transmit power over
all sub-channels in a distributed manner, while maintaining its
total data rate above a given threshold $\hat R_i$. This problem
is stated by
\begin{eqnarray}\label{TPCmulticarrier}
 \min_{\textbf{p}_i \geq 0} ~ \sum \limits_{l} p_i^l \\\nonumber
& & \hspace{-3.1cm}
 \text{subject to:} ~ \sum \log(1+\frac{{G_{i,i}^l}{p_i^l}}{\sum\limits_{j \neq
    i}{{G_{i,j}^l}{p_j^l}}+\eta_i^l}) \geq \hat R_i,
\end{eqnarray}
where $p_i^l$ is the transmit power of user $i$ over sub-channel
$l$, and $G_{i,j}^l$ is the channel gain from the transmitter of
user $i$ to the receiver of user $j$ on sub-channel $l$.

In this game, each user tries to choose an optimal power vector
$\textbf{p}_i=[p_i^1,\cdots,p_i^L]^\text{T}$, where $L$ is the
number of sub-channels that are utilized to satisfy the data rate
constraint, such that the data rate constraint in
(\ref{TPCmulticarrier}) is satisfied with the equality. Similar to
TPC, in this game, a user with a bad channel increases its
transmit power to satisfy its rate constraint, which can cause
more interference to other users, resulting in higher transmit
power levels. To make this algorithm opportunistic, we use a new
constraint and reformulate the objective function. The basic idea
is similar to OPC, i.e., users increase their transmit power in
good sub-channels, and do the opposite in bad sub-channels, but in
a more profound manner. This will lead to a higher total
throughput and a lower transmit power. We first define the utility
of each user $i$ as its total power over all sub-channels. In
opportunistic power control for multi-carrier interference
channels, each user chooses a strategy from its strategy space
that maximizes its utility. In other words, each user consumes
more power to achieve a higher data rate. As each user attempts to
maximize its utility (its total power consumed over all
sub-channels), we impose a constraint on each user's transmit
power, and provide it with incentives to behave opportunistically.
One choice for the constraint would be
\begin{equation}\label{OPCconstraint1}
    \sum \limits_{l} (p_i^l I_i^l) \leq \hat \varsigma_i,
\end{equation}
where $\hat \varsigma_i$ is a predefined upper bound for $\sum
\limits_{l} (p_i^l I_i^l)$, and
\begin{equation}\label{rieffl}
    I_i^l= \frac {\sum\limits_{j \neq i}{{G_{i,j}^l}{p_j^l}}+\eta_i^l}{G_{i,i}^l}=\sum\limits_{j \neq i} \hat G_{i,j}^l p_j^l+ \hat
    \eta_i^l.
\end{equation}
Note that, the value of $I_i^l$ depends on power levels of users
(i.e., $I_i^l(\textbf{p})$, but for simplicity in notation, we use
$I_i^l$.

Considering (\ref{OPCconstraint1}), one can observe that users
would allocate more power on good sub-channels. In addition, the
total power consumed by users depends on sub-channel conditions.
With this constraint and the objective functions as the total
power over all sub-channels, each user needs to solve a linear
program in which its entire transmit power is assigned only to the
best sub-channels, which causes instability in the iterative
algorithm. Moreover, when effective interference levels are the
same on some sub-channels, the problem would have numerous
solutions. This means that the convergence analysis of this
problem is very difficult, and convergence would be guaranteed in
a very restrictive set of channel conditions. Due to the above
problems, instead of (\ref{OPCconstraint1}), we define a new
constraint for each user as
\begin{equation}\label{OPCconstraint}
    \sum \limits_{l} (p_i^l I_i^l)^2 \leq \varsigma_i,
\end{equation}
where $\varsigma_i$ is a predefined upper bound for $\sum
\limits_{l} (p_i^l I_i^l)^2$, and $I_i^l$ is defined in
(\ref{rieffl}).

Similar to (\ref{OPCconstraint1}), applying the constraint
(\ref{OPCconstraint}) would cause each user to transmit at higher
power levels on good sub-channels, and the total transmit power
will depend on the channel conditions of sub-channels. The
strategy space of each user is
\begin{equation}\label{strategyspace}
     \mathcal{P}_i(\textbf{p}_{-i})=\{ \textbf{p}_i: p_i^l \geq 0,  \sum \limits_{l} (p_i^l I_i^l)^2 \leq
\varsigma_i  \},
\end{equation}
where $\textbf{p}_{-i}$ is the strategies of all users other than
user $i$.

The problem is formulated by the game
$\mathcal{G}^\text{o}=\langle \mathcal{M},
\mathcal{P}_i(\textbf{p}_{-i}), \{u_i\} \rangle$, where
$\mathcal{M}$ is the set of users in the system,
$\mathcal{P}_i(\textbf{p}_{-i})$ is the strategy space of user $i$
defined in (\ref{strategyspace}), and $u_i$ is the utility of user
$i$. Each user aims to solve
\begin{eqnarray}\label{OPCmulticarrier}
 \max_{\textbf{p}_i \geq 0} ~ \sum \limits_{l} p_i^l \\\nonumber
& & \hspace{-3.1cm}
 \text{subject to:} ~ \textbf{p}_i \in \mathcal{P}_i(\textbf{p}_{-i}),
\end{eqnarray}
where $\mathcal{P}_i(\textbf{p}_{-i})$ is defined in
(\ref{strategyspace}).

\subsection{Game Analysis}\label{solution analysis}

In the opportunistic power control game, each user solves
(\ref{OPCmulticarrier}) in a distributed manner. Since the
strategy space of user $i$, i.e.,
$\mathcal{P}_i(\textbf{p}_{-i})$, depends on the strategies of
other users, this game belongs to the generalized Nash equilibrium
(GNE) problems \cite{facchinei1}. In such games, in addition to
the utility function, users' interactions affect their strategy
choices. However, this dependence makes the problem hard to
analyze. In what follows, we present an analysis of
(\ref{OPCmulticarrier}). When users iteratively solve
(\ref{OPCmulticarrier}), the game settles at a GNE as defined
below.

\textbf{Definition 1}: The strategy vector
$\textbf{p}=[\textbf{p}_1^\text{*T},\cdots,\textbf{p}_M^\text{*T}]^\text{T}$
is a GNE for the game $\mathcal{G}^\text{o}$ if
\begin{equation}\label{nash}
      \sum \limits_{l} p_i^l \leq \sum \limits_{l} p_i^{l*}, ~ ~\qquad
      \forall \textbf{p}_i \in \mathcal{P}_i(\textbf{p}_{-i}^{*}),~ \forall i.
\end{equation}

When the strategies of other users are fixed, user $i$ solves the
optimization problem (\ref{OPCmulticarrier}), which is a convex
optimization. To do so, we consider its Lagrangian given by
\begin{equation}\label{lagrangefunction}
      L_i=\sum \limits_{l} p_i^l - \lambda_i (\sum \limits_{l} (p_i^l I_i^l)^2 - \varsigma_i).
\end{equation}
For a fixed but arbitrary non-negative $\textbf{p}_{-i}$, we take
the derivative of Lagrangian $L_i$ with respect to $p_i^l$, and
write
\begin{equation}\label{lagrangefunction}
      \frac{\partial L_i}{\partial p_i^l}= 1 - 2 \lambda_i p_i^l (I_i^l)^2=0.
\end{equation}
Note that at the optimal point, the constraint
(\ref{OPCconstraint}) is satisfied with the equality. Hence, the
solution to (\ref{OPCmulticarrier}) is
\begin{equation}\label{optimizationsolution}
      p_i^l=\frac{1}{2 \lambda_i (I_i^l)^2},
\end{equation}
where $\lambda_i$ is obtained such that (\ref{OPCconstraint}) is
satisfied with the equality.

Comparing (\ref{optimizationsolution}) with (\ref{OPC}), one can
see their similarity. When the effective interference experienced
by user $i$ on sub-channel $l$ is high, i.e., when the
interference from other users is high and/or the direct channel
gain from the transmitter of user $i$ to its corresponding
receiver is low, user $i$ consumes less power on that sub-channel,
and when the sub-channel is good, user $i$ increases its transmit
power. The corresponding data rate on that sub-channel is $R_i^l=
\log {(1+\gamma_i^l)}$, where $\gamma_i^l$ is the SINR of user $i$
on sub-channel $l$. If all sub-channels for user $i$ are bad, its
total transmit power is reduced, which helps users with good
sub-channels to transmit at higher rates. This means that in our
proposed scheme, a user opportunistically benefits from the
reduced transmit power levels of other users who experience bad
sub-channels.

In the game $\mathcal{G}^\text{o}$, each user updates its transmit
power over all sub-channels via (\ref{optimizationsolution}) in a
distributed and iterative manner. If the power updates converge,
they will converge to a GNE of the game. Since the strategy of a
user depends on other users' strategies, we cannot use the
existing analysis in the literature that are developed for
conventional games. Instead, we use the following theorem to prove
that the proposed game always has at least one GNE.

\textbf{Theorem 1} \cite{facchinei1}: Let
$\mathcal{G}^\text{GNEP}=\langle \mathcal{M},
\mathcal{S}_i(\textbf{s}_{-i}), \{u_i\} \rangle$ be given, where
$\mathcal{M}$ is the set of players,
$\mathcal{S}_i(\textbf{s}_{-i})$ is the strategy set of player $i$
that depends on the strategies of other users, i.e., on
$\textbf{s}_{-i}$, and $u_i$ is the utility function of user $i$.
Suppose that
\begin{list}{\labelitemi}{\leftmargin=1em}
\item[a)] There exist $M$ nonempty, convex and compact sets
$\mathcal{K}_i \subset \mathbb{R}^{L}$ such that for every
$\textbf{s} \in \mathbb{R}^{ML}$ with $\textbf{s}_i \in
\mathcal{K}_i$ for every $i$, the set
$\mathcal{S}_i(\textbf{s}_{-i})$ is nonempty, closed, and convex,
$\mathcal{S}_i(\textbf{s}_{-i}) \subseteq \mathcal{K}_i$, and
$\mathcal{S}_i$, as a point-to-set map, is both upper and lower
semi-continuous; \item[b)] For every player $i$, the function
$u_i(\cdot, \textbf{s}_{-i})$ is quasi-concave on
$\mathcal{S}_i(\textbf{s}_{-i})$, which is required in our case,
as each user tries to maximize its own utility.
\end{list}
Then a GNE exists for $\mathcal{G}^\text{GNEP}$.

From Theorem 1, we have the following theorem for the existence of
GNE in $\mathcal{G}^\text{o}$.

\textbf{Theorem 2}: The game $\mathcal{G}^\text{o}$ always admits
at least one GNE.

\textbf{Proof}: Consider the set $\mathcal{P}_i(\textbf{p}_{-i})$
as defined in (\ref{strategyspace}). Since each $\hat \eta_i^l$ in
(\ref{rieffl}) is positive, we define the set
$\mathcal{K}_i=\{\textbf{p}_i: 0 \leq p_i^l \leq \kappa_i^l \}$,
where $\kappa_i^l= \frac{\sqrt{\varsigma_i}}{\hat \eta_i^l}$. One
can easily see that the assumptions of Theorem 1 are satisfied,
i.e., the game $\mathcal{G}^\text{o}$ always has at least one GNE.
~$\blacksquare$

Although Theorem 2 states that a GNE always exists for
$\mathcal{G}^\text{o}$, it nevertheless may not be unique.
However, when GNE is not unique, the distributed algorithm may not
converge to a GNE, and may toggle between two GNEs. Below, in
Theorem 3 we provide the sufficient conditions for GNE's
uniqueness.

\textbf{Theorem 3}: The GNE in the game $\mathcal{G}^\text{o}$ is
unique if matrix $\textbf{A}$ defined as
\begin{equation}\label{uniquenessmatrix1}
 [\textbf{A}]_{i,j}= \left\{ \begin{array}{ll}
\frac{1}{\sqrt{\varsigma_i}}\min_l(\sqrt{\underline{q}_i^l} \hat \eta_i^l) & \mbox{\text{if} $i = j$},\\
-3  \sqrt{\varsigma_i}  \max_l \frac{\hat G_{i,j}^l}{\hat
\eta_i^l} & \mbox{\text{if} $i \neq j$},
\end{array} \right.
\end{equation}
is a P-matrix\footnote{A matrix $\textbf{A} \in \mathbb{R}^{n
\times n}$ is called a P-matrix if all of its principal minors are
positive \cite{pang1}.}, where
$\underline{q}_i^l=(\underline{p}_i^l \underline{I}_i^l)^2
-\sigma$, $\sigma$ is a small positive constant,
$\underline{I}_i^l=\hat \eta_i^l$, and $\underline{p}_i^l$ is the
minimum power level of user $i$ on sub-channel $l$.

\textbf{Proof}: See Appendix \ref{proof3}. ~$\blacksquare$

We also provide another sufficient condition for GNE's uniqueness
in Theorem 4 below.

\textbf{Theorem 4}: The GNE in the game $\mathcal{G}^\text{o}$ is
unique if
\begin{equation}\label{uniqunessspectral}
      \rho (\textbf{B}) <1,
\end{equation}
where $\rho (\textbf{B})$ is the spectral radius\footnote{The
spectral radius of matrix $\textbf{B}$ is its maximum absolute
eigenvalue.} of $\textbf{B}$, and
\begin{equation}\label{uniquenessmatrix2}
 [\textbf{B}]_{i,j}= \left\{ \begin{array}{ll}
~ 0 & \mbox{\text{if} $i = j$},\\
3 ~ {\varsigma_i}\frac{  \max_l \frac{\hat G_{i,j}^l}{\hat
\eta_i^l}}{\min_l(\sqrt{\underline{q}_i^l} \hat \eta_i^l)} &
\mbox{\text{if} $i \neq j$},
\end{array} \right.
\end{equation}

\textbf{Proof}: See Appendix \ref{proof4}. ~$\blacksquare$

When users in the network solve (\ref{OPCmulticarrier}), each user
chooses its transmit power iteratively in a distributed manner,
and simultaneously with other users according to
(\ref{optimizationsolution}), i.e., $p_i^l(n+1)=P(\textbf{p}(n))$,
where $P(\cdot)=\frac{1}{2 \lambda_i (I_i^l)^2}$ and
$\textbf{p}(n)$ is the transmit power levels of users at iteration
$n$. In the following theorem, we provide conditions for
convergence of this distributed algorithm.

\textbf{Theorem 5}: The distributed iterative power update
function converges to the unique GNE of the game under the same
condition as in Theorem 3.

\textbf{Proof}: See Appendix \ref{proof5}. ~$\blacksquare$

As stated earlier, global convergence of the distributed
algorithm is guaranteed if its GNE is unique. Therefore, it is not surprising that the conditions for convergence of the algorithm
are the same as those of uniqueness of its GNE.

\section{Pricing for Opportunistic Power Control}\label{opcpricing}

In game theoretic distributed schemes, each user selfishly chooses a strategy for optimizing its utility. However, this may cause
unacceptable consequences for other users, but can be controlled
via pricing in the game. Pricing is set in such a way to attain
certain desirable characteristics, such as introducing
opportunistic behavior in our case for rate maximization under
total power constraint that depends on interference levels in
sub-channels. We denote the game with no pricing by
$\mathcal{G}^\text{r}=\langle \mathcal{M}, \mathcal{P}_i, \{u_i\}
\rangle$, where $\mathcal{M}$ is the set of users, $\mathcal{P}_i$
is the strategy space of user $i$ defined by
$\mathcal{P}_i=\{\textbf{p}_i: \textbf{p}_i \geq \textbf{0}, \sum
\limits_{l} p_i^l < P_i \}$, and $u_i$ is the total throughput of
user $i$ over all sub-channels. Since the strategy space of each
user is independent of other users, this is the conventional game
where each user aims to solve
\begin{eqnarray}\label{ratemaximizationo}
     \hspace{1.2cm}  \max_{\textbf{p}_i \geq \textbf{0} } ~ \sum \limits_{l} \log(1+\frac {G_{i,i}^l p_i^l}{\sum\limits_{j \neq
      i}{{G_{i,j}^l}{p_j^l}}+\eta_i^l}) \\\label{ratemaximizationc}
    & & \hspace{-6.6cm}  \text{Subject to:} ~~ \sum \limits_{l} p_i^l \leq P_i. \nonumber
\end{eqnarray}
The solution to (\ref{ratemaximizationo}) is the well known
waterfilling, given by
\begin{equation}\label{waterfilling1}
      p_i^l=\left [\mu_i-\frac {\sum\limits_{j \neq
      i}{{G_{i,j}^l}{p_j^l}}+\eta_i^l}{G_{i,i}^l} \right]^+  \forall l,
\end{equation}
where $[\cdot]=\max(\cdot,0)$, and $\mu_i$ is chosen such that the
constraint in (\ref{ratemaximizationo}) is satisfied with the
equality.

The game $\mathcal{G}^\text{r}$ has been extensively studied in
the literature, and conditions for the uniqueness of its NE and
for the convergence of the distributed algorithm are provided in
\cite{pang2,scutari2,scutari3,scutari4}. However, as stated
earlier, due to the distributed nature of optimization and selfish
behavior of users, the output of the game may not be a desirable
one. Therefore, if users' behavior is controlled, it may be
possible to achieve certain improvements in utilizing resources
such as higher throughputs and lower power levels. To this end, we
propose a pricing mechanism that takes into account the transmit
power levels of users as well as the interference they experience.
The proposed pricing is defined by
\begin{equation}\label{pricing1}
      C(\textbf{p})=\lambda_i \sum\limits_l p_i^l I_i^l,
\end{equation}
where $I_i^l$ is the effective interference experienced by user
$i$ on sub-channel $l$ as defined in (\ref{rieffl}), and
$\lambda_i$ is the pricing for user $i$. When pricing
(\ref{pricing1}) is applied, each user is priced more when its
transmit power and/or its effective received interference are
increased. Thus, users would allocate more power on sub-channels
whose interference levels are low.

When pricing (\ref{pricing1}) is applied to the data rate
maximization problem, each user in the game
$\mathcal{G}^\text{p}=\langle \mathcal{M}, \mathcal{P}_i, \{u_i\}
\rangle$ aims to solve
\begin{eqnarray}\label{ratemaximizationpo}
     \hspace{1.2cm}  \max_{\textbf{p}_i \geq \textbf{0} } ~ \sum \limits_{l} \log(1+\frac { p_i^l}{\sum\limits_{j \neq
      i}{{\hat G_{i,j}^l}{p_j^l}}+\hat \eta_i^l})- \lambda_i \sum\limits_l p_i^l I_i^l \\\label{ratemaximizationpc}
    & & \hspace{-8.6cm}  \text{Subject to:} ~~ \sum \limits_{l} p_i^l \leq P_i. \nonumber
\end{eqnarray}
The solution to (\ref{ratemaximizationpo}), and the existence of
NE for the game $\mathcal{G}^\text{p}$ are provided in the
following theorem.

\textbf{Theorem 6}: The game $\mathcal{G}^\text{p}$ always admits
at least one NE. Moreover, each user chooses its transmit power
over each sub-channel according to
\begin{equation}\label{waterfillingpricing}
      p_i^l=\left [\frac{1}{\mu_i+\lambda_i I_i^l}-I_i^l \right]^+ ,
\end{equation}
where $I_i^l$ is defined in (\ref{rieffl}), and $\mu_i$ is so
chosen to satisfy the constraint (\ref{ratemaximizationpo}).

\textbf{Proof}: The strategy space of users are non-empty, compact
and convex subset of $L$-dimensional Euclidian space. The utility
function of each user is a continuous function of the power vector
$\textbf{p}$, and is quasi-concave function of users' power levels
$\textbf{p}_i$. Therefore, the game $\mathcal{G}^\text{p}$ always
has at least one NE. The solution to the optimization problem
(\ref{ratemaximizationpo}) which is (\ref{waterfillingpricing})
can be readily obtained using the KKT conditions for
(\ref{ratemaximizationpo}). ~$\blacksquare$

In the game $\mathcal{G}^\text{p}$, each user updates its transmit
power in a distributed and iterative manner by using
(\ref{waterfillingpricing}), where the water level
$\frac{1}{\mu_i+\lambda_i I_i^l}$ depends on the interference that
users receive in their sub-channels. Since  $\mu_i$ has the same
value for all sub-channels of user $i$, it is clear that the water
levels in those sub-channels whose interference is higher than
those of others is lower, meaning that each user assigns a smaller
power level or no power to those sub-channels. The following
theorem provides the conditions for NE's uniqueness.

\textbf{Theorem 7}: The NE of $\mathcal{G}^{p}$ is unique if the
matrix $\textbf{D}$ defined below is a P-matrix.
\begin{equation}\label{uniquenessmatrixpricing1}
 [\textbf{D}]_{i,j}= \left\{ \begin{array}{ll}
~ 1 & \mbox{\text{if} $i = j$},\\
-\max_l \left ( \frac{{\hat G_{i,j}^l}\left(1+\lambda_i
{\overline{\psi}_i^l}^2
\right){\overline{\psi}_j^l}}{{\underline{\psi}_i^l}} \right ) &
\mbox{\text{if} $i \neq j$},
\end{array} \right.
\end{equation}
where $\underline{\psi}_i^l=\hat \eta_i^l$ and
$\overline{\psi}_i^l=\sum\limits_{j }{{\hat G_{i,j}^l}{P_j}}+\hat
\eta_i^l$ and $P_j$ is defined in (\ref{ratemaximizationpo}).

\textbf{Proof}: See Appendix \ref{proof7}.  ~$\blacksquare$

Users in $\mathcal{G}^\text{p}$ update their power levels in a
distributed and iterative manner according to
(\ref{waterfillingpricing}). In the following theorem, we provide
the sufficient condition for convergence of the distributed
algorithm.

\textbf{Theorem 8}: Suppose that the matrix $\textbf{D}$ in
(\ref{uniquenessmatrixpricing1}) is a P-matrix. When users update
their power levels simultaneously according to
(\ref{waterfillingpricing}), the distributed algorithm converges
to the unique NE of the game.

\textbf{Proof}: See Appendix \ref{proof8}.  ~$\blacksquare$

\section{Simulation Results}\label{simulation}

We now present simulation results for the proposed opportunistic
power control game as well as the pricing mechanism. The system
under study is the uplink of a multi-carrier network consisting of
one base station and 5 users.

\begin{figure}[t]
  \begin{center}
    \includegraphics[width=9 cm , height=6 cm]{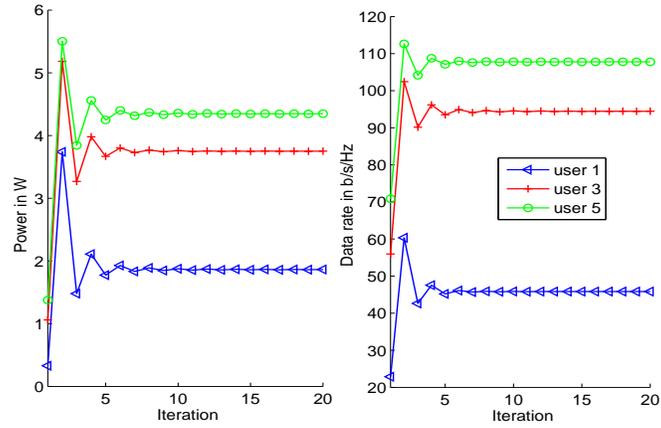} 
    \caption{Data rates and power levels of users in our proposed opportunistic power control algorithm. Channel conditions for user $i+1$ is better than that of user $i$.} 
    \label{fig:Power_rate_iteration_OPC}
  \end{center}
\end{figure}

For the opportunistic power control, the system has 20
sub-channels. The interfering channel gain from the transmitter of
user $j$ to the receiver of user $i$ on sub-channel $l$ is chosen
randomly from $(0,\frac{0.1}{i})$, where $i$ denotes the user
number, and the normalized noise power is set to 0.01 Watts.
Accordingly, the channel conditions become better from user 1 to
user 5. For an instance of the network realization, we run the
proposed algorithm and show the results in
Fig.~\ref{fig:Power_rate_iteration_OPC}. Note that user 1 whose
sub-channels are bad consumes less power and achieves a low data
rate, and user 5 with the best sub-channels transmits with high
power and achieves a high data rate.

\begin{figure}[t]
  \begin{center}
    \includegraphics[width=9 cm , height=6 cm]{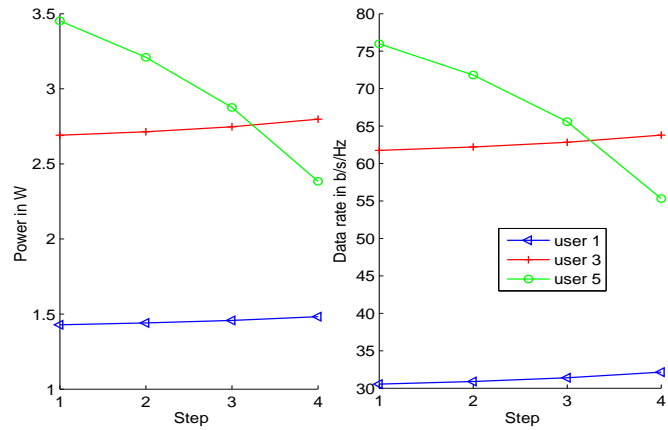} 
    \caption{Data rates and power levels of users in our proposed opportunistic power control algorithm when channel conditions for user 5 gradually worsens.} 
    \label{fig:Power_rate_step_OPC}
  \end{center}
\end{figure}

Next, we fix the channel conditions of all users except user 5 for
which in four steps, we gradually deteriorate its channel
conditions. The results are shown in
Fig.~\ref{fig:Power_rate_step_OPC}. As expected, user 5 decreases
its transmit power levels, and as the conditions for other users
improve, they increase their transmit power levels.

\begin{figure}[t]
  \begin{center}
    \includegraphics[width=9 cm , height=6 cm]{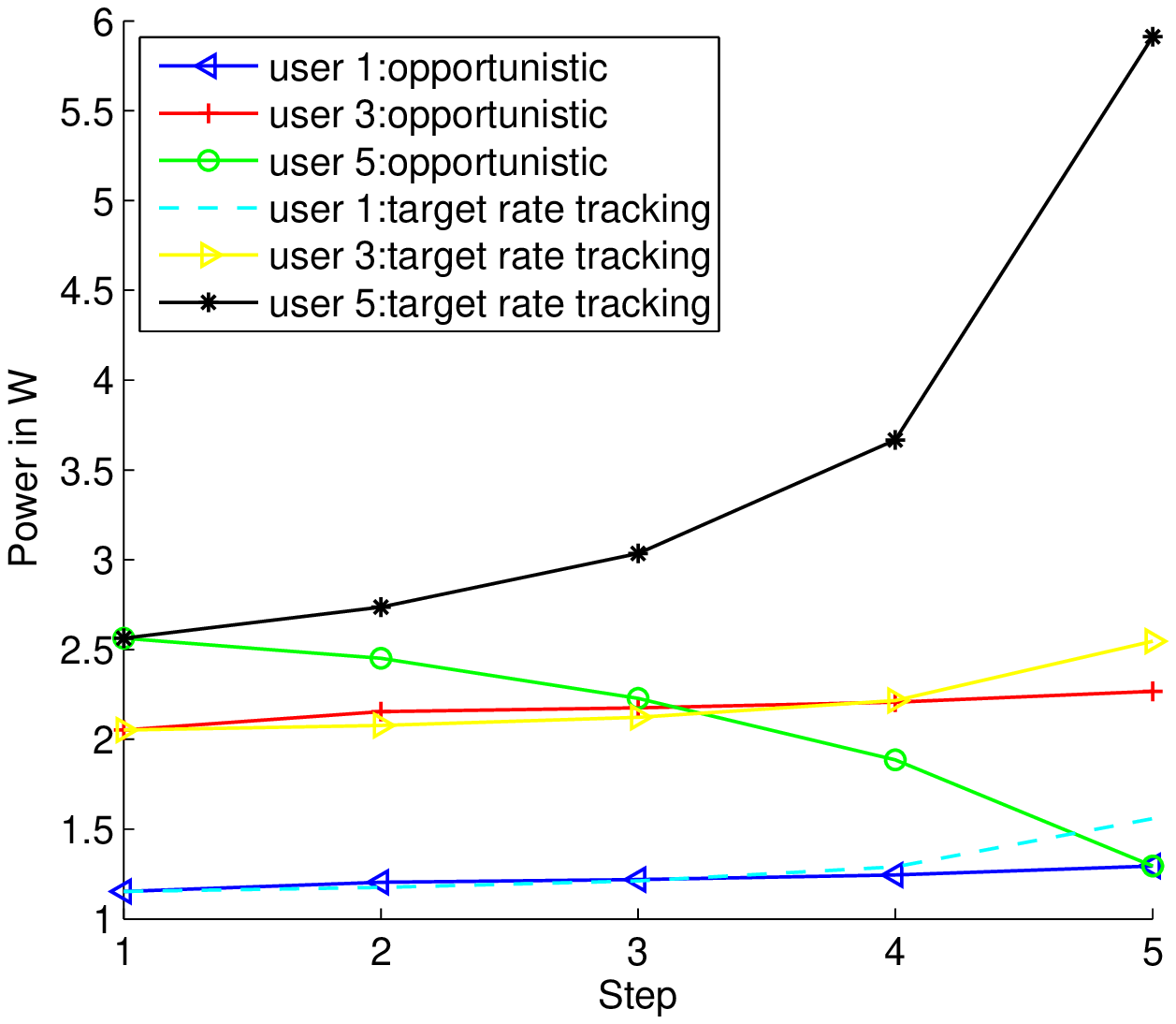} 
    \caption{Comparison of power levels in our algorithm and those in (\ref{TPCmulticarrier}) when channel conditions for user 5 gradually worsens.} 
    \label{fig:OPC_TPC_comparison}
  \end{center}
\end{figure}

As stated earlier, the behavior of the proposed algorithm is in
the opposite direction of tracking a target data rate as in
(\ref{TPCmulticarrier}). We repeat our simulations and compare the
results to those of (\ref{TPCmulticarrier}) in
Fig.~\ref{fig:OPC_TPC_comparison}. As can be seen, in
(\ref{TPCmulticarrier}), as the channel condition of user 5
deteriorates, its transmit power is increased to achieve its
target data rate. This is in contrast to our proposed algorithm,
where this user decreases its transmit power to reduce its
interference to other users. Note that, in our proposed
opportunistic scheme, at step 4, there is a 37 percent reduction
in the transmit power as compared to that of the power
minimization game (\ref{TPCmulticarrier}), but the total data rate
is reduced by 9 percent, which shows the efficiency of our
proposed scheme. In addition, using (\ref{TPCmulticarrier}) may
cause the system to become infeasible, whereas this would not
happen in our proposed game.

\begin{figure}[t]
  \begin{center}
    \includegraphics[width=9 cm , height=6 cm]{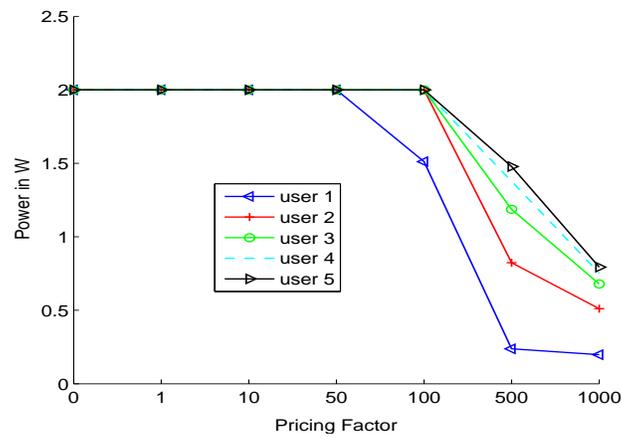} 
    \caption{The effect of pricing on the converged power levels.} 
    \label{fig:power_pricing_opportunistic_lamda_variable}
  \end{center}
\end{figure}

In the sequel, we present simulation results for the proposed
pricing. The network setup is similar to the previous simulations
except that now we have 10 sub-channels. First we show the effect
of pricing on the users' transmit power levels in
Fig.~\ref{fig:power_pricing_opportunistic_lamda_variable}. Note
that pricing affects those users with bad channels more than other
users, and forces such users to reduce their transmit power levels
more than those users with good channels, as we desired.

\begin{figure}[t]
  \begin{center}
    \includegraphics[width=9 cm , height=6 cm]{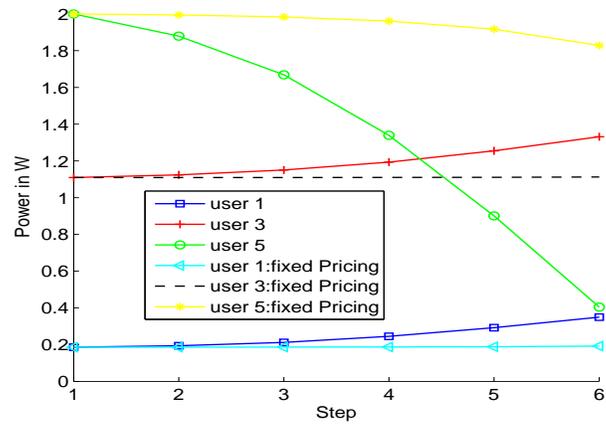} 
    \caption{Comparison of converged values of power levels using fixed pricing and those of our proposed pricing when channel conditions for user 5 gradually worsens.} 
    \label{fig:power_pricing_P_and_PI_good_bad}
  \end{center}
\end{figure}

\begin{figure}[t]
  \begin{center}
    \includegraphics[width=9 cm , height=6 cm]{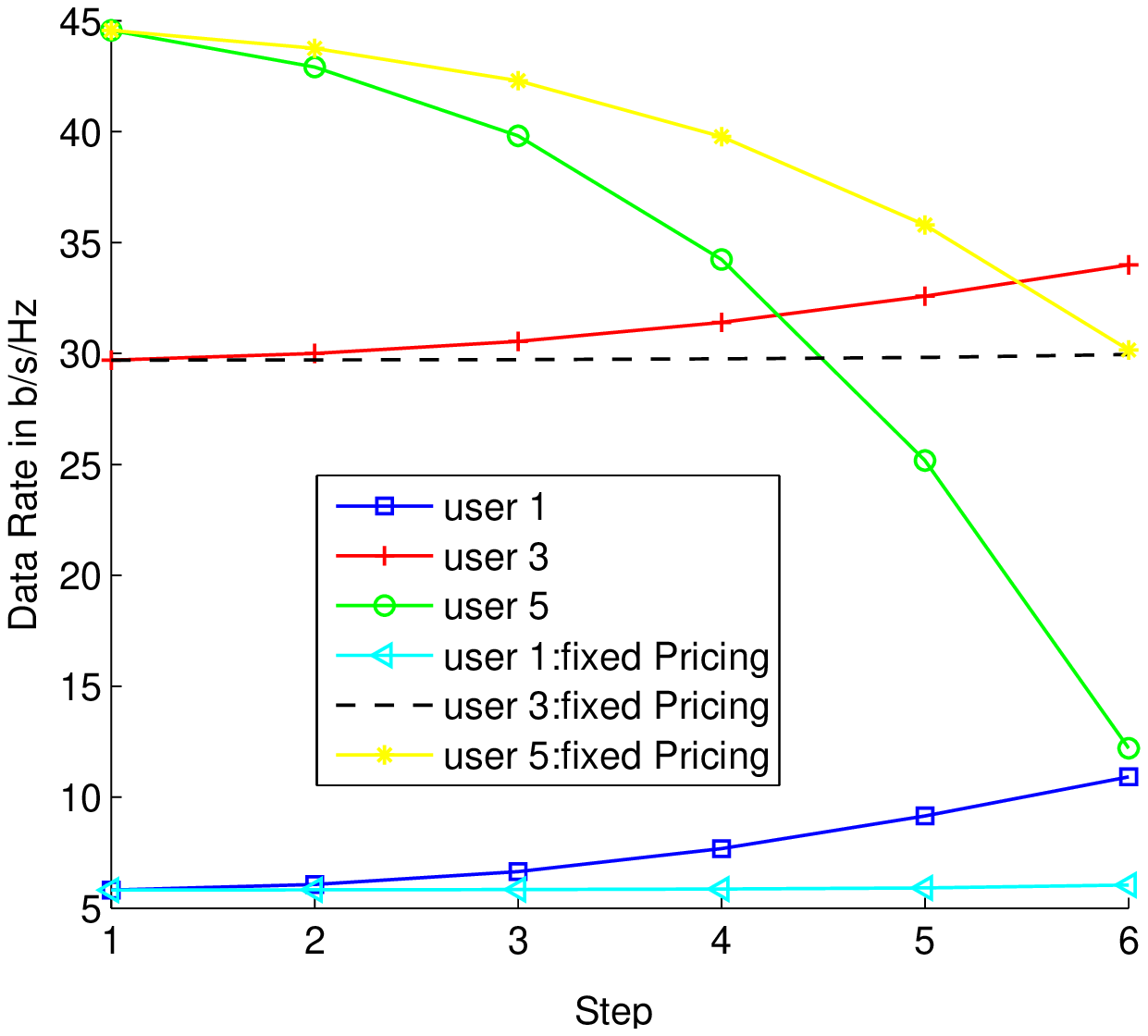} 
    \caption{Comparison of converged data rates using fixed pricing and those of our proposed pricing when channel conditions for user 5 gradually worsens.} 
    \label{fig:rate_pricing_P_and_PI_good_bad}
  \end{center}
\end{figure}

In the next two simulations, we compare the effects of our
proposed pricing with those of fixed pricing used in
\cite{scutari5}. We first run the algorithm with our proposed
pricing, and after convergence, use the multiplication of pricing
and the effective interference over each sub-channel as the fixed
pricing in \cite{scutari5}. We then use the fixed pricing for
subsequent steps in which the channel for user 5 gradually
deteriorates. The power levels are shown in
Fig.~\ref{fig:power_pricing_P_and_PI_good_bad}, and data rates in
Fig.~\ref{fig:rate_pricing_P_and_PI_good_bad}. Note that, at step
6 for our proposed pricing, the total data rates of users is about
3 percent less than that of the fixed pricing, but the total
transmit power levels is about 15 percent less than that of the
fixed pricing.

\begin{figure}[t]
  \begin{center}
    \includegraphics[width=9 cm , height=6 cm]{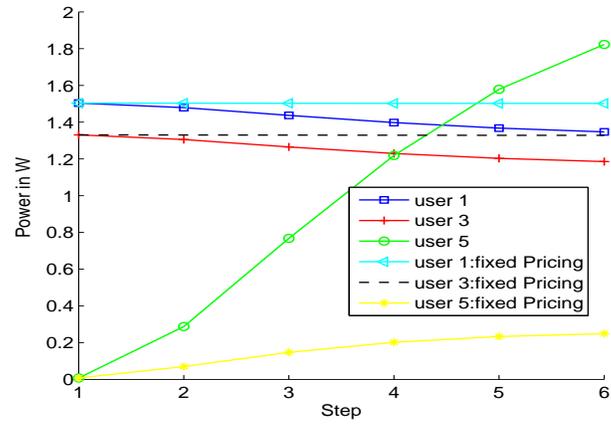} 
    \caption{Comparison of converged power levels using fixed pricing and those of our proposed pricing when channel conditions for user 5 gradually improves.} 
    \label{fig:power_pricing_P_and_PI_bad_good}
  \end{center}
\end{figure}

\begin{figure}[t]
  \begin{center}
    \includegraphics[width=9 cm , height=6 cm]{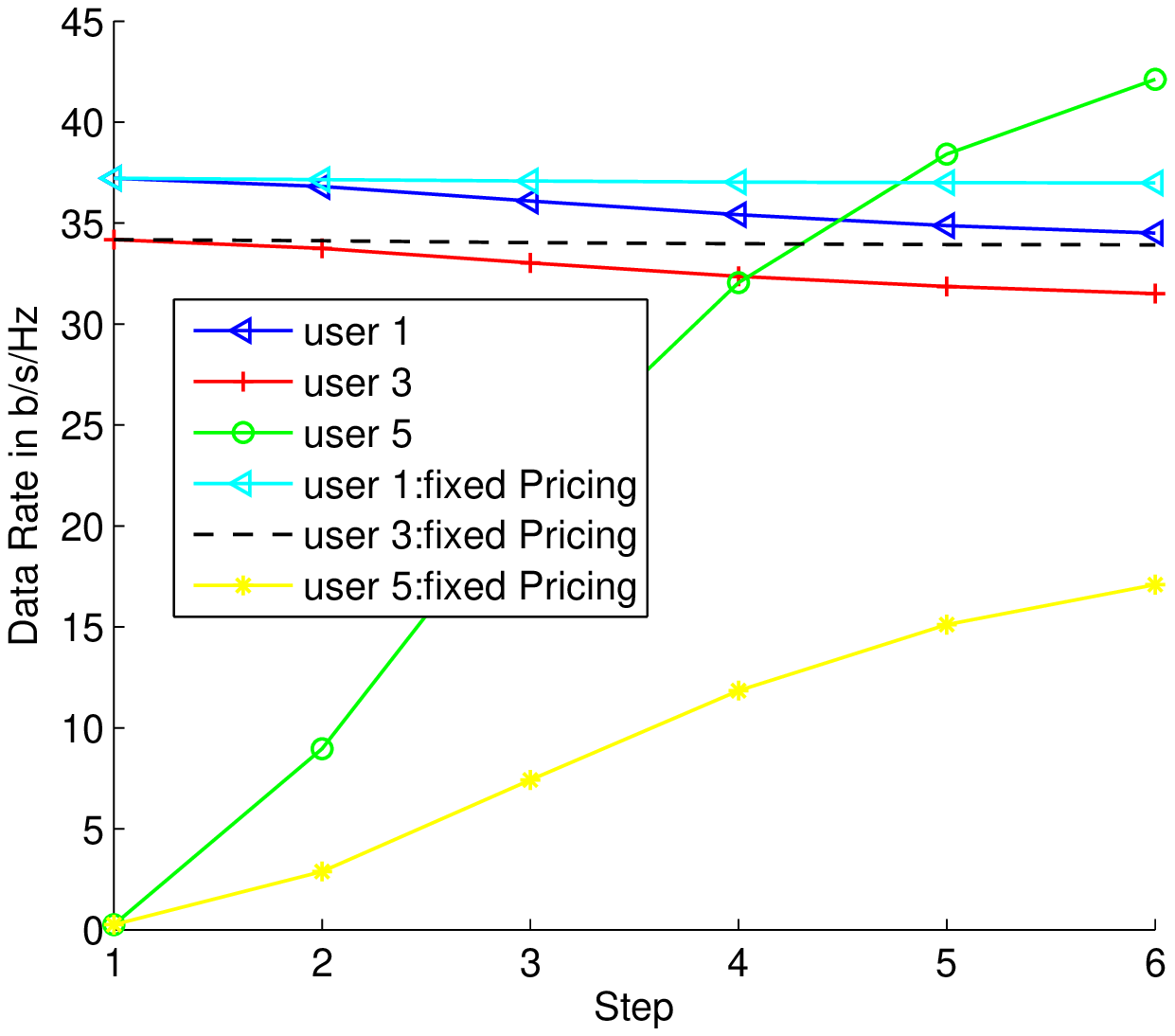} 
    \caption{Comparison of converged data rates using fixed pricing and those of our proposed pricing when channel conditions for user 5 gradually improves.} 
    \label{fig:rate_pricing_P_and_PI_bad_good}
  \end{center}
\end{figure}

Next simulations reverse the previous one, meaning that the
channels for user 5 are gradually improved at successive steps.
The power levels are shown in
Fig.~\ref{fig:power_pricing_P_and_PI_bad_good}, and data rates in
Fig.~\ref{fig:rate_pricing_P_and_PI_bad_good}. Note that at step
6, the total data rates of users as well as the total transmit
power levels are about 10 percent higher as compared to those of
the fixed pricing.

\section{Conclusions}\label{conclusion}
We proposed an opportunistic power control for multi-carrier
systems, in which each sub-channel is shared among all users. In
such a power control framework, each user transmits at lower power
levels on bad sub-channels, and does the opposite on good
sub-channels. We showed that in the proposed game there always
exists a generalized Nash equilibrium, and provided the sufficient
conditions for GNE's uniqueness and for convergence of the
distributed algorithm. Furthermore, we proposed a pricing
mechanism for the data rate maximization problem when the total
transmit power of each user is constrained depending on the
interference levels on sub-channels. In such cases, we also
provided the sufficient conditions for GNE's uniqueness, and for
convergence of the distributed algorithm. By way of simulations,
we demonstrated the improved performances of our proposed schemes
as compared to those of existing algorithms.

%

\appendices

\section{Proof of Theorem 3} \label{proof3}
We utilize variational inequalities to prove the uniqueness of NE
for the game (\ref{OPCmulticarrier}). To do so, we use the
following definition for variational inequalities.

\textbf{Definition 2} \cite{pang3}: The variational inequality
denoted by $VI(\mathcal{K},\textbf{F})$, where $\mathcal{K}$ is a
subset of $\mathbb{R}^{n}$ and $\textbf{F}:\mathcal{K} \rightarrow
\mathbb{R}^n$, is to find a vector $\textbf{x} \in \mathcal{K}$
such that $(\textbf{y}-\textbf{x})^\text{T}\textbf{F}(\textbf{x})
\geq 0$ for all $\textbf{y} \in \mathcal{K}$.

For the opportunistic power control game (\ref{OPCmulticarrier}),
the strategy space of each user depends on the strategies chosen
by other users. Therefore, we cannot directly apply the
variational inequality formulation to the game, and some
reformulations are needed. Let $\mu_i^l$ be the multiplier
corresponding to the nonnegativity constraint and $\lambda_i$ the
multiplier of the power constraint (\ref{OPCconstraint}). The KKT
conditions of the optimization problem (\ref{OPCmulticarrier}) are
\begin{eqnarray}\label{kktderivative}
     \hspace{3 cm}  - 1 + 2 \lambda_i p_i^l (I_i^l)^2 - \mu_i^l = 0, ~~ \forall l,i,\\\label{kktnonegativity}
    & & \hspace{-5.7cm}  \mu_i^l \geq 0 ~ \bot ~ p_i^l \geq 0,  ~~ \forall l,i,\\\label{kktpowerconstraint}
     \hspace{-6cm} \lambda_i \geq 0  ~ \bot ~ \sum \limits_l ({p_i^l I_i^l})^2 \leq \varsigma_i, ~~ \forall i,
\end{eqnarray}
where $\textbf{a} \bot \textbf{b}$ means that vectors $\textbf{a}$
and $\textbf{b}$ are perpendicular. Note that $\lambda_i > 0$,
otherwise the condition (\ref{kktderivative}) will lead to
$\mu_i^l < 0$, which contradicts (\ref{kktnonegativity}). This
means that the constraint $\sum \limits_l {p_i^l I_i^l}^2 \leq
\varsigma_i$ is satisfied with equality. By eliminating the
multipliers $\mu_i^l$, the KKT conditions can be reformulated as a
nonlinear complementarity problem
\begin{eqnarray}\label{kktnonegativityr}
     \hspace{3 cm}  p_i^l \geq 0 ~ \bot ~ - 1 + 2 \lambda_i p_i^l (I_i^l)^2 \geq 0, ~~ \forall l,i,\\\label{kktpowerconstraintr}
    & &  \hspace{-7 cm} \lambda_i \geq 0 ~\bot ~ \sum \limits_l ({p_i^l I_i^l})^2 - \varsigma_i=0, ~~ \forall i.
\end{eqnarray}
We define the variable $q_i^l$ by
\begin{equation}\label{alphail}
      q_i^l=({p_i^l I_i^l})^2,
\end{equation}
and use it to reformulate (\ref{OPCconstraint}) as
\begin{equation}\label{opcconstraintchangevariable}
      \sum \limits_l q_i^l=\varsigma_i.
\end{equation}

Note that $q_i^l=0$ if and only if $p_i^l=0$. On the other hand,
for each value of $\mathbf{q}^l=[q_i^l]_{i=1}^{M}$, the
corresponding values of $p_i^l$ can be obtained using the OPC
algorithm. Hence, we write $p_i^l=\theta_i^l (\mathbf{q}^l)$.
Since $\theta_i^l (\cdot)$ is a continuous function of
$\mathbf{q}^l$ \cite{sung1}, and considering $p_i^l$ and $I_i^l$
as functions of $\mathbf{q}^l$, we reformulate the conditions
(\ref{kktnonegativityr}) and (\ref{kktpowerconstraintr}) by
\begin{eqnarray}\label{kktnonegativityrr}
     \hspace{3 cm}  q_i^l \geq 0   ~ \bot ~ - 1 + 2 \lambda_i p_i^l (I_i^l)^2 \geq 0, ~~ \forall l,i,\\\label{kktpowerconstraintrr}
    & &  \hspace{-7 cm} \lambda_i \geq 0  ~ \bot ~ \sum \limits_l q_i^l = \varsigma_i, ~~ \forall i,.
\end{eqnarray}
Note that the variable $q_i^l$ is nonnegative, i.e., $q_i^l > 0$.
One can see from (\ref{optimizationsolution}) that $p_i^l$ is
always positive, i.e., $p_i^l > 0$, and since $I_i^l > 0$, we have
$q_i^l > 0$. The maximum power level for each user on each
sub-channel is $\overline{p}_i^l=\frac{\sqrt{\varsigma_i}}{\hat
\eta_i^l}$.

Now suppose that all users except user $i$ transmit at their
maximum power levels only on sub-channel $l$. The minimum power
level of user $i$ on sub-channel $l$, denoted by
$\underline{p}_i^l$, is obtained from
(\ref{optimizationsolution}). Therefore, $q_i^l >
\underline{q}_i^l$, where $\underline{q}_i^l=(\underline{p}_i^l
\underline{I}_i^l)^2 -\sigma$ and $\sigma$ is a small positive
constant, and $\underline{I}_i^l=\hat \eta_i^l$. From the above,
we change the conditions (\ref{kktnonegativityrr}) and
(\ref{kktpowerconstraintrr}) as
\begin{eqnarray}\label{kktnonegativityrrr}
     \hspace{3 cm}  q_i^l -\underline{q}_i^l \geq 0  ~ \bot ~ - 1 + 2 \lambda_i p_i^l (I_i^l)^2 \geq 0, ~~ \forall l,i,\\\label{kktpowerconstraintrrr}
    & &  \hspace{-7 cm} \lambda_i \geq 0  ~ \bot ~ \sum \limits_l q_i^l = \varsigma_i, ~~ \forall  i.
\end{eqnarray}
The conditions in (\ref{kktnonegativityrrr}) and
(\ref{kktpowerconstraintrrr}) are not equivalent to
(\ref{kktnonegativityrr}) and (\ref{kktpowerconstraintrr}).
However, all solutions to (\ref{kktnonegativityrr}) and
(\ref{kktpowerconstraintrr}) are also solutions to
(\ref{kktnonegativityrrr}) and (\ref{kktpowerconstraintrrr}). The
change in (\ref{kktnonegativityrr}) may yield additional solutions
to (\ref{kktnonegativityrrr}) and (\ref{kktpowerconstraintrrr}).
Thus, solutions to (\ref{kktnonegativityrrr}) and
(\ref{kktpowerconstraintrrr}) consist of all solutions to
(\ref{kktnonegativityrr}) and (\ref{kktpowerconstraintrr}) plus
possible other solutions. Note that $\lambda_i > 0$. We further
reformulate (\ref{kktnonegativityrrr}) and
(\ref{kktpowerconstraintrrr}) into a more suitable form as
\begin{eqnarray}\label{kktvariationalnonegativity}
     \hspace{3 cm}  q_i^l -\underline{q}_i^l \geq 0  ~ \bot ~ \xi_i - \log(p_i^l)+ \log(q_i^l)\geq 0, ~~ \forall l,i, \\\label{kktvariationalconstraint}
    & &  \hspace{-8 cm}  \xi_i \in \mathbb{R}, ~  ~ \sum \limits_l q_i^l = \varsigma_i, ~~ \forall i.
\end{eqnarray}

We use log transform because we wish to eliminate the
multiplication of the power $p_i^l$ and the effective interference
$I_i^l$. It is obvious that (\ref{kktvariationalnonegativity}) and
(\ref{kktvariationalconstraint}) are the KKT conditions of the
variational inequality $VI(\mathcal{X},\textbf{F})$ where
$\mathcal{X}=\prod \limits_i \mathcal{X}_i$ and
$\mathcal{X}_i=\{\mathbf{q}_i \in \mathbb{R}^L: q_i^l \geq
\underline{q}_i^l, \sum \limits_l q_i^l = \varsigma_i \}$, and
$F_i^l=\log(q_i^l)- \log (p_i^l)$. With this modification, we now
provide conditions for GNE's uniqueness. Note that all GNEs of the
game are solutions to $VI(\mathcal{X},\textbf{F})$. However, the
variational inequality may have additional solutions. Hence, the
conditions for uniqueness of the solution to
$VI(\mathcal{X},\textbf{F})$ guarantee GNE's uniqueness as well.
This condition, however, may be excessive, since solutions to
$VI(\mathcal{X},\textbf{F})$ may not be the GNE of the game.

Let $\widehat{\textbf{q}}= \mathbf{q}(1)$ and
$\widetilde{\textbf{q}}=\mathbf{q}(2)$ be two solutions to
$VI(\mathcal{X},\textbf{F})$. This means that for each user $i$,
we have
\begin{equation}\label{variational1}
      \sum \limits_l (q_i^l(2) - q_i^l(1))\left(- \log ({p}_i^l(1))+ \log (q_i^l(1)) \right)\geq 0,
\end{equation}
\begin{equation}\label{variational2}
      \sum \limits_l (q_i^l(1) - q_i^l(2)) \left(- \log ({p}_i^l(2))+ \log (q_i^l(2)) \right) \geq 0.
\end{equation}
In addition, from the definition of $q_i^l$, we have
\begin{equation}\label{differenceoftwosolutions}
      \log (q_i^l(2))-\log(q_i^l(1))=2 \left(\log ({p}_i^l(2))-\log ({p}_i^l(1)) + \log({I}_i^l(2))-\log({I}_i^l(1)) \right).
\end{equation}
We add (\ref{variational1}) and (\ref{variational2}), and write
\begin{equation}\label{variational3}
      \sum \limits_l (q_i^l(1) - q_i^l(2)) \left( \log ({p}_i^l(2)) - \log({p}_i^l(1))+ 2 \log ({I}_i^l(2))-2 \log ({I}_i^l(1)) \right) \geq 0.
\end{equation}
From the mean value theorem, we know that there exists a $q_i^l(1)
\leq q_i^l \leq q_i^l(2)$ such that
\begin{equation}\label{meanvaluealpha}
     q_i^l(1) - q_i^l(2)=q_i^l
 \left( \log (q_i^l(1))- \log (q_i^l(2)) \right).
\end{equation}
Therefore, using (\ref{differenceoftwosolutions}) and
(\ref{meanvaluealpha}), (\ref{variational3}) can be written as
\begin{eqnarray}\nonumber
    \hspace{1 cm} \sum \limits_l q_i^l \left( \log ({p}_i^l(1))-\log ({p}_i^l(2)) + \log ({I}_i^l(1))-\log ({I}_i^l(2) \right) \\\label{variational4}
    & &  \hspace{-7 cm}  \left( \log ({p}_i^l(2))-\log ({p}_i^l(1))+ 2 \log ({I}_i^l(2))-2 \log ({I}_i^l(1)) \right) \geq 0.
\end{eqnarray}
Rearranging (\ref{variational4}) and using Schwarz's inequality,
we get
\begin{equation}\label{variational5}
     \sqrt{ \sum \limits_l q_i^l \left( \log ({p}_i^l(1))-\log ({p}_i^l(2)) \right)^2} \leq 3 \sqrt{\sum \limits_l q_i^l
     \left( \log ({I}_i^l(1))-\log ({I}_i^l(2)) \right)^2},
\end{equation}
or equivalently
\begin{equation}\label{variational6}
     \| \sqrt{\mathbf{q}} (\log ({\textbf{p}}(1))-\log ({\textbf{p}}(2))) \| \leq 3  \| \sqrt{\mathbf{q}}(\log ({\textbf{I}}(1))-\log ({\textbf{I}}(2))) \|,
\end{equation}
where $\|\cdot\|$ denotes the Euclidian norm. Again, we use the
mean value theorem for $\log ({p}_i^l)$ and $\log ({I}_i^l)$, and
write
\begin{equation}\label{variational7}
     \sqrt{ \sum \limits_l q_i^l (\frac{{p}_i^l(1) - {p}_i^l(2)}{p_i^l})^2} \leq
     3\sqrt{\sum \limits_l q_i^l (\frac{{I}_i^l(1) - {I}_i^l(2)}{I_i^l})^2},
\end{equation}
and
\begin{equation}\label{variational7}
     \sqrt{ \sum \limits_l q_i^l (\frac{{p}_i^l(1) - {p}_i^l(2)}{p_i^l})^2} \leq
     3\sqrt{\sum \limits_l q_i^l (\frac{1}{I_i^l} \sum \limits_{j\neq i} \hat G_{i,j}^{l}({p}_j^l(1) -
     {p}_j^l(2)))^2}.
\end{equation}
Note that $p_i^l \leq \frac{\sqrt{\varsigma_i}}{\hat \eta_i^l}$,
$I_i^l \geq \hat \eta_i^l $, and $\underline{q}_i^l \leq q_i^l
\leq \varsigma_i$. From (\ref{variational7}), we obtain
\begin{equation}\label{variational7}
     \frac{1}{\sqrt{\varsigma_i}}\min_l(\sqrt{\underline{q}_i^l} \hat \eta_i^l)\sqrt{ \sum \limits_l ({p}_i^l(1) - {p}_i^l(2))^2} \leq
     3 \sum \limits_{j\neq i} \sqrt{\varsigma_i}  \max_l \frac{\hat G_{i,j}^l}{\hat \eta_i^l} \sqrt{\sum \limits_{l}({p}_j^l(1) -
     {p}_j^l(2))^2}.
\end{equation}
Defining $[\textbf{a}]_i=\sqrt{ \sum \limits_l (p_i^l(1) -
p_i^l(2))^2}$ and considering the matrix $\textbf{A}$ defined in
(\ref{uniquenessmatrix1}), we obtain $\textbf{A} \textbf{a} \leq
\textbf{0}$. Therefore, if the matrix $\textbf{A}$ is a P-matrix,
we have $\textbf{a}=\textbf{0}$, and hence the proof.


\section{Proof of Theorem 4}\label{proof4}
From (\ref{variational7}), one obtains
\begin{equation}\label{variational8}
     \sqrt{ \sum \limits_l ({p}_i^l(1) - {p}_i^l(2))^2} \leq
     3 \sum \limits_{j\neq i} \frac{\sqrt{\varsigma_i}  \max_l \frac{\hat G_{i,j}^l}{\hat \eta_i^l}}{ \frac{1}{\sqrt{\varsigma_i}}\min_l(\sqrt{\underline{q}_i^l} \hat \eta_i^l)} \sqrt{(\sum \limits_{l}({p}_j^l(1) - {p}_j^l(2)))^2},
\end{equation}
If the matrix
\begin{equation}\label{uniquenessmatrixintermediate1}
 [\widehat{\textbf{{B}}}]_{i,j}= \left\{ \begin{array}{ll}
~ 1 & \mbox{\text{if} $i = j$},\\
-3  \frac{\sqrt{\varsigma_i}  \max_l \frac{\hat G_{i,j}^l}{\hat
\eta_i^l}}{
\frac{1}{\sqrt{\varsigma_i}}\min_l(\sqrt{\underline{q}_i^l} \hat
\eta_i^l)}  & \mbox{\text{if} $i \neq j$},
\end{array} \right.
\end{equation}
is a P-matrix, GNE is unique. From the P-property of the matrix,
this is equivalent to the spectral condition in
(\ref{uniqunessspectral}).

\section{Proof of Theorem 5}\label{proof5}
Note that at each iteration, say $n+1$, the parameters $q_i^l$ and
power vectors are related by
\begin{equation}\label{variational8}
    q_i^l (n+1)= (p_i^l(n+1) I_i^l(\textbf{p}(n))^2.
\end{equation}
Since they are the solutions of the game (\ref{OPCmulticarrier}),
we have
\begin{eqnarray}\label{kktvariationalnonegativityiterative}
     \hspace{3 cm}  q_i^l(n+1) -\underline{q}_i^l \geq 0,  ~ ~ \xi_i^{n+1} - \log (p_i^l(n+1))+ \log (q_i^l(n+1))\geq
     0,
     \\\label{kktvariationalconstraintiterative}
    & &  \hspace{-10 cm}  \xi_i^{n+1} \in \mathbb{R}, ~ ~ \sum \limits_l q_i^l(n+1) = \varsigma_i.
\end{eqnarray}
Therefore, with ${p}_i^l(1)=p_i^l(n+1)$ and ${p}_i^l(2)=p_i^{l*}$,
where $p_i^{l*}$ is the GNE of the game, one can follow the same
line as in proof of Theorem 3 to obtain the following inequality
\begin{equation}\label{variationaliterative}
      \frac{1}{\sqrt{\varsigma_i}}\min_l(\sqrt{\underline{q}_i^l} \hat \eta_i^l) \sqrt{ \sum \limits_l (p_i^l(n+1) - p_i^{l*})^2} \leq
     3 \sum \limits_{j\neq i} \sqrt{\varsigma_i}  \max_l \frac{\hat G_{i,j}^l}{\hat \eta_i^l} \sqrt{(\sum \limits_{l}(p_j^l(n+1) -
     p_j^{l*}))^2}.
\end{equation}
Using (\ref{variationaliterative}) and the P-property of matrix
$\textbf{A}$ defined in (\ref{uniquenessmatrix1}), one can easily
derive the condition.

\section{Proof of Theorem 7}\label{proof7}
Let $\mu_i$ be the multiplier corresponding to the power
constraint (\ref{ratemaximizationpc}). The KKT conditions for the
optimization problem (\ref{ratemaximizationpo}) can be
reformulated as the following complementarity problem
\begin{eqnarray}\label{kktvariationalpricingnonnegativity}
     \hspace{1.2cm}  p_i^l \geq 0 ~ \bot ~ -(\frac {1}{\sum\limits_{j }{{\hat G_{i,j}^l}{p_j^l}}+\hat \eta_i^l})+ \lambda_i  I_i^l+\mu_i \geq 0, ~~ \forall l,i, \\\label{kktvariationalpricingpowerc}
    & & \hspace{-8.6cm} \mu_i \geq 0 ~ \bot ~  \sum \limits_{l} p_i^l \leq P_i, ~~ \forall i.
\end{eqnarray}
These are the KKT conditions for $VI(\mathcal{X},\textbf{F})$,
where $\mathcal{X}=\prod \limits_i \mathcal{X}_i$ and
$\mathcal{X}_i=\{\textbf{p}_i \in \mathbb{R}^L: p_i^l \geq 0, \sum
\limits_lp_i^l \leq P_i \}$, and
\begin{equation}\label{variationalfunctionpricing}
    F_i^l(\textbf{p})=-(\frac {1}{\sum\limits_{j }{{\hat G_{i,j}^l}{p_j^l}}+\hat \eta_i^l})+ \lambda_i
    I_i^l.
\end{equation}
Since each set $\mathcal{X}_i$ is closed and convex, and
$F_i^l(\cdot)$ is continuous, $VI(\mathcal{X},\textbf{F})$ has a
solution. Since the set $\mathcal{X}$ is a Cartesian product of
some independent closed and convex sets, it is known \cite{pang3}
that $VI(\mathcal{X},\textbf{F})$ has a unique solution if
$\textbf{F}(\cdot)$ is uniformly P-function, which means that
there exists a constant $c$ such that for every $\textbf{p} \in
\mathcal{X}$ and $\textbf{p}' \in \mathcal{X}$, we have
\begin{equation}\label{uniformlypcondition}
    \max_i (\textbf{p}_i-\textbf{p}_i')(\textbf{F}_i(\textbf{p})-\textbf{F}_i(\textbf{p})) \geq c
    \|\textbf{p}-\textbf{p}' \|^2.
\end{equation}
Therefore, it suffices to prove that the function $F(\cdot)$ is
uniformly P-function. For $\textbf{F}(\cdot)$, we have
\begin{equation}\label{variationalfunctionpricingdifference}
  F_i^l(\textbf{p})-F_i^l(\textbf{p}'))= \frac{\sum\limits_{j
  }{{\hat G_{i,j}^l}({p_j^l}-{p_j^{l}} ' )}}{(\sum\limits_{j }{{\hat G_{i,j}^l}{{p_j^l}}}+\hat \eta_i^l)(\sum\limits_{j
  }{{\hat G_{i,j}^l}{{p_j^l}}'}+\hat \eta_i^l)}+ \lambda_i \sum\limits_{j
  \neq i
  }{{\hat G_{i,j}^l}({p_j^l}-{p_j^{l}} ' )}.
\end{equation}
We define the following variables
\begin{equation}\label{saifunction}
    \psi_i^l=\sqrt{(\sum\limits_{j }{{\hat G_{i,j}^l}{{p_j^l}}}+\hat \eta_i^l)(\sum\limits_{j
  }{{\hat G_{i,j}^l}{{p_j^l}}'}+\hat \eta_i^l)},
\end{equation}
and write
\begin{eqnarray}\label{uniformpfunctionproof1}
     \hspace{0.1cm}  (\textbf{p}_i-\textbf{p}_i')(\textbf{F}_i(\textbf{p})-\textbf{F}_i(\textbf{p}))= \sum \limits_{l}(p_i^l-{p_i^l}')\frac{\sum\limits_{j}{{\hat G_{i,j}^l}({p_j^l}-{p_j^{l}} ' )}}{(\sum\limits_{j
}{{\hat G_{i,j}^l}{{p_j^l}}}+\hat \eta_i^l)(\sum\limits_{j
  }{{\hat G_{i,j}^l}{{p_j^l}}'}+\hat \eta_i^l)}\\\label{uniformpfunctionproof2}
    & & \hspace{-13.6cm} + \lambda_i \sum\limits_{j
  \neq i
  }{{\hat G_{i,j}^l}({p_j^l}-{p_j^{l}} ' )}= \sum \limits_{l}\frac{(p_i^l-{p_i^l}')^2}{{\psi_i^l}^2} + \sum \limits_{l}\frac{(p_i^l-{p_i^l}') \sum\limits_{j \neq i}{{\hat G_{i,j}^l}\left(1+\lambda_i {\psi_i^l}^2\right)({p_j^l}-{p_j^{l}} ') }}{{\psi_i^l}^2}
  \\\label{uniformpfunctionproof3}
   & & \hspace{-13.6cm} \geq \sum \limits_{l}\frac{(p_i^l-{p_i^l}')^2}{{\psi_i^l}^2} - \sum\limits_{j \neq i} \left | \sum \limits_{l} (p_i^l-{p_i^l}')(p_j^l-{p_j^l}')\frac {{\hat G_{i,j}^l}
   \left(1+\lambda_i {\psi_i^l}^2 \right)}{{\psi_i^l}^2}\right | \\\label{uniformpfunctionproof5}
   & & \hspace{-13.6cm} \geq \sum \limits_{l}\frac{(p_i^l-{p_i^l}')^2}{{\psi_i^l}^2} - \sum \limits_{j \neq i} \left(\max_l \left ( \frac{{\hat G_{i,j}^l}\left(1+\lambda_i {\psi_i^l}^2 \right){\psi_j^l}}{{\psi_i^l}} \right )\sqrt{ \sum
\limits_{l} \frac{(p_i^l-{p_i^l}')^2}{{\psi_i^l}^2}} \sqrt{ \sum
\limits_{l} \frac{(p_j^l-{p_j^l}')^2}{{\psi_j^l}^2}} \right),
\end{eqnarray}
where we applied Schwartz' inequality to
(\ref{uniformpfunctionproof3}). By some manipulations, one can
obtain the following inequality
\begin{equation}\label{uniformpfunctionproof6}
    (\textbf{p}_i-\textbf{p}_i')(\textbf{F}_i(\textbf{p})-\textbf{F}_i(\textbf{p}))
    \geq [\textbf{d}]_i [\textbf{D}\textbf{d}]_i,
\end{equation}
where $\textbf{d}$ is a vector whose $i^\text{th}$ element is
$[\textbf{d}]_i=\sqrt{ \sum \limits_{l}
\frac{(p_i^l-{p_i^l}')^2}{{\psi_i^l}^2}}$ and $\textbf{D}$ is
defined in (\ref{uniquenessmatrixpricing1}). From
(\ref{uniformpfunctionproof6}) and the P-property assumption on
matrix $\textbf{D}$, one can show that $\textbf{F}(\cdot)$ is
uniformly P-function \cite{pang1,scutari5}.

\section{Proof of Theorem 8}\label{proof8}
To prove the convergence of the algorithm, we follow the same line
as in the proof of Theorem 5. Given the power profile
$\textbf{p}(n)$ at iteration $n$, users update their power levels
according to (\ref{waterfillingpricing}). This means that their
power levels $p_i^l(n+1)$ must satisfy the following optimality
condition
\begin{equation}\label{optimality algorithmconvergep1}
    \sum \limits_{l} (p_i^l-p_i^l(n+1))(-\frac{1}{\sum\limits_{j \neq i} {{\hat G_{i,j}^l}{{p_j^l(n)}}}+\hat \eta_i^l}+ \lambda
    I_i^l(p(n))).
\end{equation}
The NE must also satisfy a similar condition, i.e.,
\begin{equation}\label{optimality algorithmconvergepn}
    \sum \limits_{l} (p_i^l-{p_i^l}^*)(-\frac{1}{\sum\limits_{j \neq i} {{\hat G_{i,j}^l}{{p_j^l}^*}}+\hat \eta_i^l}+ \lambda
    I_i^l(p^*)).
\end{equation}
Adding these two inequalities and following the same steps as in
Theorem 5, this theorem is proved.


\begin{thebibliography}{10}
\providecommand{\url}[1]{#1} \csname url@rmstyle\endcsname
\providecommand{\newblock}{\relax}
\providecommand{\bibinfo}[2]{#2}
\providecommand\BIBentrySTDinterwordspacing{\spaceskip=0pt\relax}
\providecommand\BIBentryALTinterwordstretchfactor{4}
\providecommand\BIBentryALTinterwordspacing{\spaceskip=\fontdimen2\font
plus \BIBentryALTinterwordstretchfactor\fontdimen3\font minus
  \fontdimen4\font\relax}
\providecommand\BIBforeignlanguage[2]{{%
\expandafter\ifx\csname l@#1\endcsname\relax
\typeout{** WARNING: IEEEtran.bst: No hyphenation pattern has been}%
\typeout{** loaded for the language `#1'. Using the pattern for}%
\typeout{** the default language instead.}%
\else \language=\csname l@#1\endcsname \fi #2}}

\bibitem{fudenberg1}
D.~Fudenberg and J.~Tirole, \emph{Game Theory}.\hskip 1em plus
0.5em minus
  0.4em\relax Cambridge, MA: MIT Press, 1991.

\bibitem{han1}
Z.~Han, Z.~Ji, and K.~J.~R. Liu, ``Non-cooperative resource
competition game by
  virtual referee in multi-cell {OFDMA} networks,'' \emph{IEEE Journal on
  Selected Areas in Communications}, vol.~25, no.~6, pp. 1–--10, August 2007.

\bibitem{scutari1}
J.-S. Pang, G.~Scutari, F.~Facchinei, and C.~Wang, ``Distributed
power
  allocation with rate constraints in {G}aussian parallel interference
  channels,'' \emph{IEEE Transactions on Information Theory}, vol.~54, no.~8,
  pp. 2868–--2878, August 2008.

\bibitem{facchinei1}
F.~Facchinei and C.~Kanzow, ``Generalized {N}ash equilibrium
problems,''
  \emph{4OR}, vol.~5, pp. 173--–210, 1993.

\bibitem{foschini1}
G.~J. Foschini and Z.~Miljanic, ``A simple distributed autonomous
power control
  algorithm and its convergence,'' \emph{IEEE Transactions on Vehicular
  Technology}, vol.~42, no.~4, pp. 641--646, November 1993.

\bibitem{sung1}
K.~Leung and C.~W. Sung, ``An opportunistic power control
algorithm for
  cellular network,'' \emph{IEEE/ACM Transactions on Networking}, vol.~14,
  no.~3, pp. 470–--478, June 2006.

\bibitem{scutari2}
G.~Scutari, D.~P. Palomar, and S.~Barbarossa, ``Optimal linear
precoding
  strategies for wideband noncooperative systems based on game theory {-}
  {P}art {I}: Nash equilibria,'' \emph{IEEE Transactions on Signal Processing},
  vol.~56, no.~3, pp. 1230--–1249, March 2008.

\bibitem{scutari3}
------, ``Optimal linear precoding strategies for wideband noncooperative
  systems based on game theory {—} {P}art {II}: Algorithms,'' \emph{IEEE
  Transactions on Signal Processing}, vol.~56, no.~3, pp. 1250–--1267, March
  2008.

\bibitem{scutari4}
------, ``Asynchronous iterative water-filling for {G}aussian
  frequency-selective interference channels,'' \emph{IEEE Transactions on
  Information Theory}, vol.~54, no.~7, pp. 2868–--2878, July 2008.

\bibitem{pang1}
R.~W. Cottle, J.-S. Pang, and R.~E. Stone, \emph{The Linear
Complementarity
  Problem}.\hskip 1em plus 0.5em minus 0.4em\relax Cambridge Academic Press,
  1992.

\bibitem{sung2}
K.~W. Shum, K.-K. Leung, and C.~W. Sung, ``Convergence of
iterative
  waterfilling algorithm for {G}aussian interference channels,'' \emph{IEEE
  Journal on Selected Areas in Communications}, vol.~25, no.~6, pp. 1091--
  1100, August 2007.

\bibitem{bertsekas1}
D.~P. Bertsekas and J.~N. Tsitsiklis, \emph{Parallel and
Distributed
  Computation: Numerical Methods}, 2nd~ed.\hskip 1em plus 0.5em minus
  0.4em\relax Athena Scientific, 1989.

\bibitem{boyd1}
S.~Boyd and L.~Vandenberghe, \emph{Convex Optimization}.\hskip 1em
plus 0.5em
  minus 0.4em\relax Cambridge University Press, 2004.

\bibitem{pang2}
Z.-Q. Luo and J.-S. Pang, ``Analysis of iterative waterfilling
algorithm for
  multiuser power control in digital subscriber lines,'' \emph{EURASIP Journal
  on Applied Signal Processing}, vol. 2006, pp. 1–--10, May 2006.

\bibitem{pang3}
F.~Facchinei and J.-S. Pang, \emph{Finite-Dimensional Variational
Inequalities
  and Complementarity Problem}.\hskip 1em plus 0.5em minus 0.4em\relax
  Springer-Verlag, New York, 2003.

\bibitem{krunz1}
F.~Wang, M.~Krunz, and S.~Cui, ``Price-based spectrum management
in cognitive
  radio networks,'' \emph{IEEE Journal on Selected Topics in Signal
  Processing}, vol.~2, no.~1, p. 74 –87, February 2008.

\bibitem{scutari5}
J.-S. Pang, G.~Scutari, D.~P. Palomar, and F.~Facchinei, ``Design
of cognitive
  radio systems under temperature-interference constraints: A variational
  inequality approach,'' \emph{IEEE Transactions on Signal Processing},
  vol.~58, no.~6, pp. 3251--3271, June 2010.

\end{thebibliography}

\clearpage



\end{document}